\documentclass[10pt,letterpaper]{article}
\usepackage{caption}
\usepackage{amsmath}
\usepackage{amsfonts}
\usepackage{subcaption}

\usepackage[top=0.85in,left=1.in,footskip=0.75in,marginparwidth=2in]{geometry}

\usepackage[utf8]{inputenc}

\usepackage{cite}

\usepackage{nameref,hyperref}

\usepackage{microtype}
\DisableLigatures[f]{encoding = *, family = * }

\raggedright
\usepackage{changepage}

\usepackage[aboveskip=1pt,labelfont=bf,labelsep=period,singlelinecheck=off]{caption}

\makeatletter
\renewcommand{\@biblabel}[1]{\quad#1.}
\makeatother

\usepackage{lastpage,fancyhdr,graphicx}
\usepackage{epstopdf}
\fancyheadoffset[L]{2.25in}
\fancyfootoffset[L]{2.25in}

\usepackage{color}

\definecolor{Gray}{gray}{.25}

\usepackage{graphicx}

\usepackage{sidecap}

\usepackage{wrapfig}
\usepackage[pscoord]{eso-pic}
\usepackage[fulladjust]{marginnote}
\reversemarginpar

\begin{document}
\vspace*{0.35in}

\begin{flushleft}
{\Large
\textbf\newline{Quantifying Retail Agglomeration Using Diverse Spatial Data}
}
\newline
\\
Duccio Piovani\textsuperscript{1,*},
Vassilis Zachariadis\textsuperscript{2},
Michael Batty\textsuperscript{1},
\\
\bigskip
\bf{1} Centre for Advanced Spatial Analysis (CASA), University College London (UCL), 90 Tottenham Court Road , London, W1T 4TJ
\\
\bf{2} Prospective Labs Ltd, IDEA London, 69 Wilson Street, Shoreditch, London, EC2A 2BB
\\
\bigskip
* d.piovani@ucl.ac.uk
\end{flushleft}

\section*{Abstract}
Newly available data on the spatial distribution of retail activities in cities makes it possible to build models formalized at the level of the single retailer. 
Current models tackle consumer location choices at an aggregate level and the opportunity new data offers for modeling at the retail unit level lacks a theoretical framework. The model we present here helps to address these issues. It is a particular case of the \emph{Cross-Nested Logit} model, based on random utility theory built with the idea of quantifying the role of floor space and agglomeration in retail location choice. We test this model on the city of London: the results are consistent with a super linear scaling of a retailer's \emph{attractiveness} with its floor space, and with an agglomeration effect approximated as the total retail floor space within a $325m$ radius from each shop.

\section*{Introduction}
Our approach to understanding and studying cities has radically changed in the last two decades \cite{batty1995new,bettencourt2010unified}
. Advances in spatial network theory \cite{barthelemy2011spatial,crucitti2006centrality}, dramatic increases in the volume and type of available data \cite{batty2013big} and insightful  approaches based on out-of-equilibrium concepts  \cite{louf2013modeling,louf2014congestion}  have contributed to the formation of a new, highly interdisciplinary, \emph{science of cities}. It is common knowledge that a fundamental role in the urbanization process is played by retail activity and indeed a city's social life and physical shape is greatly influenced by such activities. There is continual motion in cities as people move to work and play, in restaurants, bars, grocery shops, shopping malls and so on. Understanding and describing the mechanisms that govern these activities and the processes that lead to their agglomeration is not only intriguing, but crucial to an understanding of how a city works. For this reason modeling retail location has been a pillar of urban simulation for a long time\cite{huff1966programmed}. Researchers have attempted  different types of approaches to characterize this phenomenon, from entropy maximizing models \cite{wilson2011entropy,wilson1969use}, to random utility models \cite{Teller2008,manchanda1999shopping, williams1977formation} to agent based approaches \cite{vanhaverbeke2011agent,Heppenstall2013}. However, retail location and consumer location choice theory continue to draw heavily from fundamental ideas \cite{von1966isolated,hotelling1990stability} in central place theory \cite{christaller1966central,dennis2002central} and rent-bid theory \cite{alonso1960theory}. Since the late 1990s, advances in the new economic geography \cite{krugman1990increasing,krugman1998s} have reinvigorated the field and reorganized it, by considering economies of scale, cross-dependencies in markets (e.g. between labour, retail and housing) and forms of imperfect competition between firms \cite{dixit1977monopolistic}. However, apart from notable exceptions \cite{anderson1992discrete,suarez2004accounting,fiasconaro2016spatio}, fresh approaches have been slow in informing state-of-the-art modelling and engaging with mainstream location choice modelling based on discrete choice and random utility theory.
Here we present such a model of consumer location choice based on random utility theory designed to capture both internal and external economies of scale at the individual retailer level, which are represented by a retailer’s size and the size of its neighbors. 
In the following discussion, we will examine the model’s theoretical foundations, and propose an implementation strategy which takes advantage of unconventional data sources that have recently become available for calibration and validation. We then review the model’s output predictions. Our study focuses on central London and the data gives us information on the spatial distribution of the population and work places (from the 2011 Population Census), on work to retail and home to retail trips (from the London Travel Demand Survey 2012) which is be used to calibrate the cost function of our model, and on the position, size and paid rent of single retailers (from the Valuation Office Agency 2010) which we use to quantify the effect of internal and external economies. Finally we use data on Foursquare check-ins to validate the results. A richer description of the datasets we use is found below in the Methods section. As we will see, this work shows the central role retail agglomeration plays in consumer choice, and how this is created by local interactions at the \emph{microscopic} level of the single retailer. In our model, consumers are evaluating the attractiveness of each shop by considering its size and the retail activity with a $325m$ perimeter around it. This is in line with observations reported in literature on the average length of walking trips.
\section*{Results}
\noindent
\textbf{The Model.} Using as input the spatial distribution of population  and individual retailer locations in London, our goal is to build a model of consumer choice capable of predicting the \emph{success} of a retailer to a high level of accuracy. This will be measured by the expected number of people that will visit the retail store, or in other words, by the fraction of the distribution of trips that the model predicts will flow towards each single retailer. 
We start by defining the utility of consumer $i$ shopping for product $p$ from retailer $r$ as
\begin{equation}
u_{ir} = u_p - p_r - c(d_{ir}) + \omega_{ir}
\label{eq:ui}
\end{equation}
where $u_p$ is the utility that comes by acquiring the product, $p_r$ the price at which the product is sold by retailer $r$, $c(d_{ir})$ a generalised cost of traveling between $r$ and $i$ and where $\omega_{ir}$ is a random element that reflects the personal tastes of the consumer for location and product variation. As one can see eq.{\ref{eq:ui}} is a random utility function. Now following \cite{train2009discrete}, if one now considers the random element $\omega_{ir}$ as i.i.d. Gumble distributed for all retailers, by integrating over it, we can define the probability of consumer $i$ shopping in $r$ as the probability that the utility is the greatest of  other alternatives, namely $P(u_{ir} > u_{ir\prime})\,\, \forall r\prime$, which has the form
\begin{equation}
p_{i\rightarrow r} = \frac{\text{exp}\,(-\beta_i u_{ir})}{\sum_{r'} \text{exp}\,(-\beta_i u_{ir'}) } =\frac{\text{exp}\,(-\beta_i (u_p-p_r-c(d_{ir})))}{\sum_{r'} \text{exp}\,(-\beta_i (u_p - p_{r'} - c(d_{ir'}))) } 
\label{eq:p1}
\end{equation}
where $\beta_i$ is the inverse of the standard deviation of the distribution of $\omega_{ir}$. If we now consider $\beta_i = \beta $ for every consumer, eq.(\ref{eq:p1}) will depend only on the location of consumer $i$ and retailer $r$. 

Eq.(\ref{eq:p1}) implies that each retailer $r$ offers one, and only one, type of product $p$. Therefore, each consumer $i$ associates only one random utility component $\omega_{ir}$ with each retailer, reflecting the fit between this sole product variation and the preferences of the consumer. This assumption is quite unrealistic; typically retailers will stock several types of products. It is reasonable to assume that larger shops will stock larger varieties. Therefore, variety can be expressed as a function of floor space. Following the work of Daly and Zachary \cite{daly1978improved} and  the work of Ben-Akiva and Lerman in \cite{ben1985discrete}, and assuming that the price is constant for every retailer in the system, namely $p_r=p=\text{cost}$, the probability of consumer $i$ shopping from retailer $r$ is 
\begin{equation}
p_{i\rightarrow r} = \frac{f_r^\alpha \text{exp}\,(- \beta\, c(d_{ir}))}{\sum_{r'}f_{r'}^\alpha   \text{exp}\,(-\beta \, c(d_{ir'}) }
\label{eq:p2}
\end{equation}
where $f_r$ is the floor space of retailer $r$ and the exponent $\alpha$ quantifies the correlation between the stochastic components and the product variation. The model in eq.(\ref{eq:p2}) is very similar to the multinomial logit model presented by McFadden \cite{mcfadden2001economic,mcfadden1980econometric} and to the entropy maximising equation introduced by Wilson \cite{wilson2011entropy}. Macroscopically, this suggests that the utility that consumer $i$ gets from buying from retailer $r$ is either a sub-linear ($\alpha < 1$), linear ($\alpha = 1$), or super-linear ($\alpha > 1$) function of the floor space of retail unit r. These internal economies (or dis-economies) of scale emerge from the application of a transparent utility-based approach and reflect perceived opportunities at the level of the individual retailer. Eq.(\ref{eq:p2}) captures the potential impact of size at the individual shop level. As such it is sufficient in describing flow patterns and activity distribution associated with shop-size variations. But as we will see in the following section, it fails to account for the concentration of retail activity in clusters (markets/shopping centers). In other words, eq.(\ref{eq:p2}) cannot explain agglomeration effects in the spatial distribution of retail activity. 

In order to explicitly introduce agglomeration into the model, we start by defining the attractiveness of a retailer as
\begin{equation}
A_r = f_r^\alpha + \sum_{r':d_{rr'}} f_{r'}^\alpha \text{exp}\,(-\gamma \,\,d_{rr'} )
\label{eq:A}
\end{equation}
where $d_{rr'}$ is the distance between retailers, and $\gamma$ is parameter that weights the role of the retailer's neighborhood in the consumer's perception. The perceived utility associated with a given retailer is influenced by the utility associated with the neighboring 
retailers. By substituting eq.(\ref{eq:A}) into eq.(\ref{eq:p2}) we finally get the form of the distribution to be used in our model as:
\begin{equation}
p_{i\rightarrow r} = \frac{A_r\text{exp}\,(- \beta\, c(d_{ir}))}{\sum_{r'}A_{r'} \text{exp}\,(-\beta \, c(d_{ir'})) }
\label{eq:Pfinal}
\end{equation}
In the location choice model presented in eq.(\ref{eq:Pfinal}), we can see how the parameter $\alpha$ controls the internal and $\gamma$ the external economies  of scale. Once again $\alpha$ tells us if the utility scales sub-, super- or linearly with the retailer's floor space, 
while the $\gamma$ values tune the \emph{interaction range}  between retailers: $\gamma \rightarrow \infty$ implies that the utility of visiting a retailer is only a function of its own floor space $f_r$, while in the opposite case, when $\gamma \rightarrow 0 $, the utility of a retailer is equally shaped by all shops in its vicinity. Note that for $\alpha = 1$ and $\gamma \rightarrow \infty $, we derive the multinomial logit model. The model in eq.(\ref{eq:Pfinal}) is a particular case of the \emph{Cross-Nested} Logit model (CNL) \cite{wen2001generalized,bierlaire2006theoretical} in which retailer r represents a nest, and the extent of the overlap between nests is controlled by $\gamma$.

Now exploiting eq.(\ref{eq:Pfinal}), we can define the modeled turnover, i.e. the fraction of consumers that will shop in a given destination, for each retailer $r$ as 
\begin{equation}
Y_r  = \sum_{i} n_i \cdot p_{i\rightarrow r} = \sum_l \left( \frac{A_r \text{e}^{- \beta C(d_{ir})}}{\sum_{r\prime}  A_{r\prime}\text{e}^{- \beta C(d_{ir\prime} )} }\right) \cdot n_i
\label{eq:Y}
\end{equation}
where we have considered the population concentrated on centroids $i$,and where $n_i$ indicates their respective population. Therefore the term $ n_i \cdot p_{i\rightarrow r}$ tells us the fraction of the population of centroid $i$ that will travel to retailer $r$, and depends both on the attractiveness $A_r$ of the retailer and on the distance between the two points. It will then be possible to  compare the results yielded by eq.(\ref{eq:Y}) with the actual rents paid by retailers. Quantifying the two parameters $\alpha$ and $\gamma$ provides us with insight into the economical relationships between retailers which is the main objective of this paper. To do this we have to first calibrate the cost function which expresses how consumers perceive costs related to the distance between their origin and a shop.  \\

\noindent
\textbf{Calibration of Cost Function.} We define a cost function $C(d,\lambda)$ as a Box-Cox transformation of the distance, namely
\begin{equation}
c(d,\lambda) = \frac{d^\lambda - 1}{\lambda}
\label{eq:cost}
\end{equation}
where $d$ is the distance between the consumer and the retailer's location. The Box-Cox function adds an extra modeling layer that maps objective costs into perceived costs. The curve of this transformation ranges from linear ($\lambda = 1$) to logarithmic  ($\lambda \rightarrow 0$). In the former case the marginal effect of cost travel $d$ on trip distribution is $\text{exp}(-\beta * d)$, while in the latter it becomes $d^{-\beta}$.
To calibrate the two parameters $\beta$ and $\lambda$ we use the London Travel Demand Survey (LTDS) database which contains 5004 home to retail  and 2242 work to retail trips (see Methods for more details). These numbers are not sufficient to calibrate the location choice model in full; we thus cannot use this dataset to give value to all the parameters used in model. In fact the LTDS supplementary report (TfL 2011) suggests that the sample size is sufficient only for an Inner/Outer London spatial classification. Therefore the survey is only used to calibrate distance profiles i.e. the distribution of distance traveled for shopping from home and work. As an input for the population distribution, we have used London Population Census data at the Lower Super Output Layer level (LSOA), where we have considered the population concentrated on the centroids. As we can see below in the Methods section, the database includes coordinates of the LSOA centroids, population and working population.

We start by arbitrarily assigning values to the $\alpha$ and $\gamma$  parameters, which sets the values for the retailers' \emph{attractiveness} in eq.(\ref{eq:A}). We chose $\alpha = \{0.8,1.0,1.8\}$ in order to cover the 3 different scenarios of sub-linear, linear and super-linear scaling of attractiveness with floor space, and $\gamma = \{0.005,0.05,0.5\}$ which bearing in mind that we are working in meters, corresponds to a decay length of $\{200m, 20m,2m\}$ respectively. 
 
\begin{figure}[t]
\centering
\begin{subfigure}[b]{0.45\textwidth}
	\includegraphics[width=\textwidth]{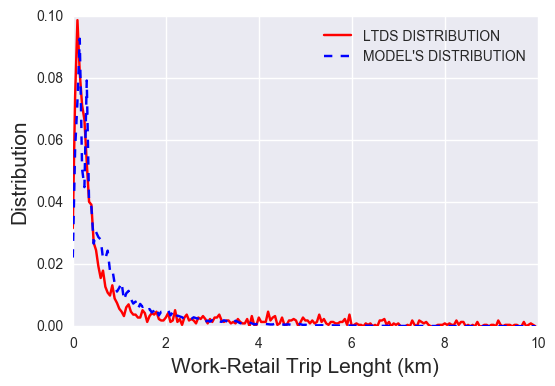}
	\caption{}
	\label{fig:W_R}
\end{subfigure}
\begin{subfigure}[b]{0.45\textwidth}
	\includegraphics[width=\textwidth]{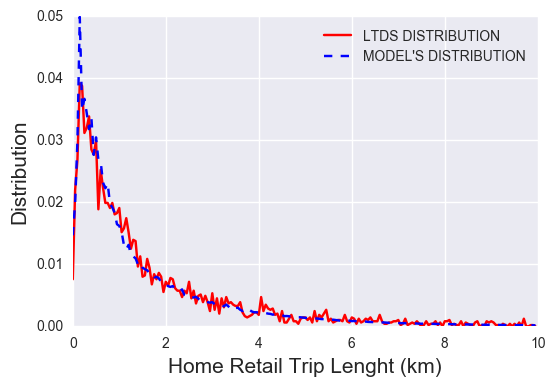}
    \caption{}
	\label{fig:H_R}
\end{subfigure}
\caption{ We show the maximum likelihood modeled distributions and the observed ones. For work-retail trips \ref{fig:W_R} we have found $\beta_w=0.35,\lambda_w=0.25$, while for home-retail in  trips \ref{fig:H_R} $\beta_h=0.25,\lambda_h = 0.3 $. As we can see from the figures in both cases the parameters that maximize the likelihood generate a distribution that fits the observed one very well.}
\label{fig:Distr}
\end{figure}

\noindent
\newline
Now exploiting eq.(\ref{eq:Pfinal}) we can define the probability the model generates a trip of distance $d$ as
\begin{equation}
p_m(d) = \frac{\sum_k n_k \sum_{r : d < d_{rk} < d + b_d} A_r \cdot e^{-\beta C(\lambda,d_{rk})}}{\sum_{k\prime} n_{k\prime} \sum_{r} A_r \cdot e^{-\beta C(\lambda,d_{r{k\prime}})}}
\label{eq:pm}
\end{equation}
where the $b_d$  in the numerator is a conveniently defined bin. In other words eq.(\ref{eq:pm}) tells us the fraction of trips that are of length $d$. To fix the $\beta$ and $\lambda$ values, we tune them so to maximize the likelihood equation
\begin{equation}
\mathbb{L}(\beta,\lambda) =  \sum_d p_e(d) \cdot\text{ln} (p_m(d))
\end{equation}
where $d$ is the length of the trips and $p_{e}(d)$ is its observed distribution in the LTDS. We repeat this process on both types of trips, in fig.(\ref{fig:Distr}) and for every couple $(\alpha,\gamma)$ in the previous list we obtain a couple $(\beta,\lambda)$ that maximizes the likelihood. The results yielded for $\beta$ and $\lambda$ are robust for the different $\alpha$, and $\gamma$ inputs, and the variations are of  $\pm 0.1$. As we can see from fig.(\ref{fig:Distr}) there is a good agreement in both cases between the distributions observed in the data and those generated by the model with parameter values as $\beta_h=0.25,\lambda_h=0.3$, $\beta_w=0.35,\lambda_w=0.25$. 
\newline
\newline
\noindent 
\textbf{Correlation with Data.} Considering the two different types of trips, the total modeled turnover predicted by the model is 
\begin{equation}
Y_r  = Y^w_r + Y^h_r = \sum_i \left( n^w_i p^w_{i\rightarrow r} +n^h_i p^h_{i\rightarrow r} \right)
\label{eq:Ytot}
\end{equation}
and at an aggregated level this quantity can be used an indicator of the \emph{success} of a retailer, with the idea that the more people visit a retailer the more that retailer earns. A more careful analysis should definitely
take into account the type of goods sold by the retailer: indeed given the same number of visits, a car show room and a grocery corner shop yield different earnings, and a clothing shop in the vicinity of another clothing shop will have a different impact to that of a supermarket. These are two significant aspects that we leave for more detailed analysis in further research. 

All the information on the retailers are found in the Valuation Office Agency database (see Methods VOA for more details), where we have found, among other details, information on the position, floor space and the ratable value per meter for 98936 retailers in London. The rateable value represents the Valuation Office Agency's estimate of the open market annual rental value of a business/non-domestic property, i.e. the rent the property would let for on the valuation date, if it were being offered on the open market; as such, it is considered a very good indicator of the property value. Our hope is that the number of \emph{visits} estimated by the model in eq.(\ref{eq:Ytot}) to a retailer of any type, could be a good predictor of its rent. To test this hypothesis we calculate the Pearson correlation between the two quantities
\begin{equation}
Y_\text{sq}(\alpha,\gamma) = \frac{Y_r(\alpha,\gamma)}{f_r} \qquad\qquad R_\text{sq}= \frac{\text{Rateable Value}}{f_r}
\label{eq:Quantities}
\end{equation}
where $f_r$ is the floor space. Not dividing by $f_r$ would result in higher levels of correlation, not generated however through higher accuracy of the model but because of the variable  $f_r$ explicitly appearing in eq.(\ref{eq:A}), and because of the strong correlation between the two quantities

\noindent
\newline
We therefore study the behavior of 
\begin{equation}
C\left( \alpha , \gamma\right) = \frac{\text{cov}\left( Y_\text{sq} , R_\text{sq}\right)}{\sigma_{Y_\text{sq} }\sigma_{R_\text{sq}}}
\label{eq:corr}
\end{equation}
looking for the values $\left( \alpha_\text{max},\gamma_\text{max}\right) $ that maximize the correlation. These parameters will be considered those that best describe the retail activity through internal and external economies. 
\begin{figure}[ht!]
\centering
\begin{subfigure}[b]{0.45\textwidth}
	\includegraphics[width=\textwidth]{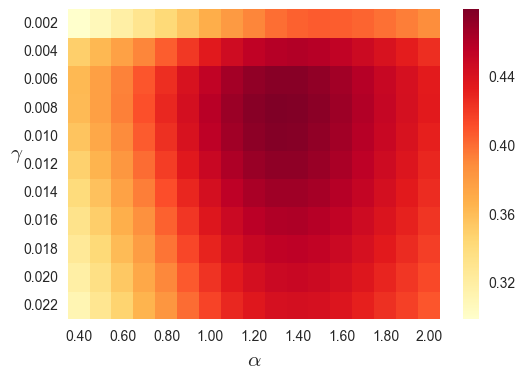}
	\caption{}
	\label{fig:H_Corr}
\end{subfigure}\quad\begin{subfigure}[b]{0.45\textwidth}
	\includegraphics[width=\textwidth]{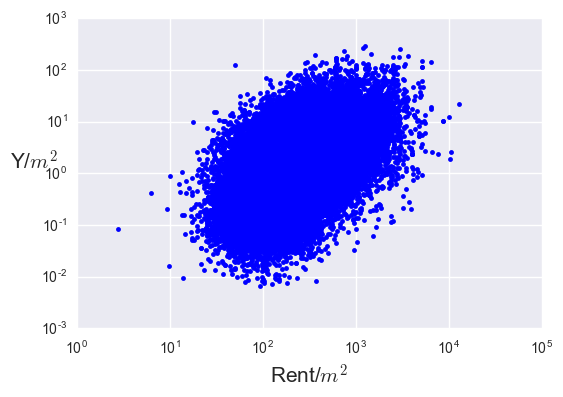}
	\caption{}
	\label{fig:Scatter}
\end{subfigure}

\begin{subfigure}[b]{0.45\textwidth}
	\includegraphics[width=\textwidth]{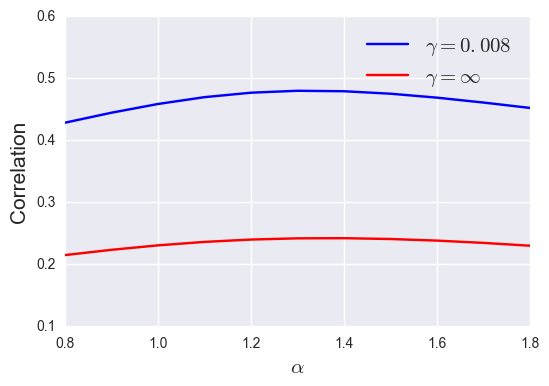}
	\caption{}
	\label{fig:s_corr}
\end{subfigure}\quad\begin{subfigure}[b]{0.45\textwidth}
	\includegraphics[width=\textwidth]{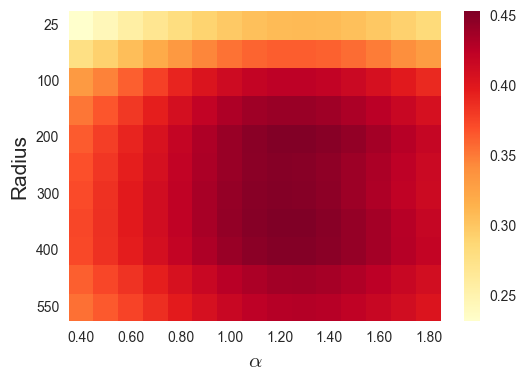}
	\caption{}
	\label{fig:r_corr}
\end{subfigure}
\label{fig:corr}
\caption{ In this figure we present the results obtained comparing the retailer's ratable value found in the VOA dataset, with the modeled turnovers $Y(\alpha,\gamma)$.In \ref{fig:s_corr} we compare the correlation in the case where $\gamma = \infty$, or with \emph{no interactions}, which is the form used in other more aggregated models. We can see how in this microscopic formalization the model always correlates better with the data for finite values of $\gamma$. In \ref{fig:H_Corr} we can see a clear maximum in the correlation for $\alpha_\text{max} = 1.3$ and $\gamma_\text{max}=0.008$, and a scatter plot of the two quantities is found in \ref{fig:Scatter}. In \ref{fig:r_corr} we show the correlation with the model of $\gamma = 0$, which implies that the attractiveness becomes $A_r = f_r^\alpha  +\sum_{r' : d_{rr'}<d_o}f_r^\alpha$, and we have studied the correlation for different values of $d_o$. We can see how the maximum is found in $\alpha_\text{max}=1.2$ and $d_\text{max}= 325m$ which is $2.6$ times the characteristic length of the decay and coincides with a dampening effect of $93\%$ of the interactions. }
\end{figure}
In fig.({\ref{fig:H_Corr}}) we show a heat map of the correlations found  exploring the parameter space. The maximum correlation is 
$C_\text{max} = 0.48$ which seems like a reasonable value given the level of aggregation we are working at and the parameter values are $\left(\alpha_\text{max},\gamma_\text{max} \right) = (1.3 , 0.008)$. As we can see the $\alpha_\text{max}$ tells us that a model with a super linear scaling between the number of visitors and floor space best describes the data, while the $\gamma_\text{max}$ value indicates that retailers benefit from their proximity to other retailers given a characteristic length of the decay of $\frac{1}{\gamma}=125m$. In fig.(\ref{fig:Scatter}) we show a scatter plot of the  two quantities in eq.(\ref{eq:Quantities}) at their maximum correlation. 

We can see in fig.({\ref{fig:H_Corr}}) how for increasing values of $\gamma$, which imply a faster decay and therefore a shorter \emph{interaction range} between retailers, the correlation levels decrease. In the extreme case of $\gamma = \infty$, the attractiveness term in eq.(\ref{eq:A}) becomes $A_r = f_r^\alpha$ meaning that retailer attractiveness only depends on its floor space. An $A_r$  of this form is what is usually found in standard retail models such as those in \cite{wilson2011entropy}, and works well when considering retail centers. From these results it seems that for a microscopical formalization of the retail activity, this picture fails as a satisfactory description. In fig.(\ref{fig:s_corr}) we compare the correlations obtained from the model with \emph{interactions} and with \emph{no interactions}. We can see how both models exhibit the maximum for the same value of $\alpha$ and how the model with interaction has constantly higher correlation values. This means that agglomeration has to be included when modeling consumers' choice at the level of the single retailer. 

Another interesting scenario is the opposite case with $\gamma = 0 $, where consumers perceive the neighborhood of a retailer in the same way  for a given distance. The attractiveness term in eq.(\ref{eq:A}) becomes 
\begin{equation}
A_r = f_r^\alpha  +\sum_{r' : d_{rr'}<d_o}f_r^\alpha
\label{eq:gzeroA}
\end{equation}
where we have introduced a parameter $d_o$ which explicitly sets the \emph{interaction range} between retailers. We can now study the Pearson correlation for different values of the radius $d_o$ and $\alpha$. In fig.(\ref{fig:r_corr}) we can see how the maximum values are found for $\alpha =1.2 $ which is very similar to what has been previously found for $d^\text{max}_o = 325 $, but despite the correlation levels being very close $C_\text{max} = 0.45$ for the model with the exponential decay, this model typically shows lower values. What this seems to imply is that consumers implicitly take into account the distance and weight closer shops more than more distant ones, instead of equally considering their proximity. 

In order to better quantify the level of agglomeration predicted by the model we can now compare the results in fig.(\ref{fig:H_Corr}) and fig.(\ref{fig:r_corr}). As implied, an exponent of $\gamma = 0.008$ indicates a characteristic decay length of $ \frac{1}{\gamma} = 125m$ for the interaction range. If we compare this result with the $\gamma = 0$ model, we see that $d^\text{max}_o = 2.6\cdot\,\frac{1}{\gamma} $, and if we substitute it in the exponential, we get $\text{e}^{- 0.008 d^\text{max}_o }  = 0.07 $ which means that at that distance retailers loose $93\%$ of the benefits of neighboring floor space. This analysis suggests that retail activity benefits from agglomeration, and this benefit is spread through local interactions between single retailers, whose main effect is felt within $325m$. From the consumer's point of view, we could say that when choosing the retailer to visit, consumers take into account the amount of floor space or choice within a radius of around $325m$. This results is in very good agreement with several studies on walking trips as in \cite{yang2012walking} where it is shown that $65\%$ of all walking trips are under $0.25 \text{miles}\simeq 400 \text{m}$, or in \cite{millward2013active} where the average walking distance for several types of retail activity are just above $\simeq 500 \text{m}$. We interpret these results as follows: the choice perceived when considering a specific retailer is quantifiable as its floor space, and the floor spaces of retailers which can be reached by foot.



\section*{Discussion}

In this section we will work through the tests we have done in order to validate the levels of agglomeration and the scaling law predicted by the model. We have done so by repeating the same procedure both on a randomized data set and on data coming from Foursquare. We have randomized the data both in the rents $/m^2$, and in their locational positions. To randomize the rents=$/m^2$, we have first measured their distribution in the VOA dataset and then assigned a random rent to each retailer taken from the same measured distribution, while to randomize the position we have simply reshuffled the position of retailers taking as lower and upper bounds the minimum and maximum coordinates found in the data. In both cases the correlation between the rents and the \emph{turnovers} predicted by the model  disappeared completely ($C_\text{max} \simeq 10^{-3}$). This seems to imply that there is a clear relation between the retailers' spatial interactions and the rents they pay, and that these are captured, at an aggregated level by eq.(\ref{eq:A}).

A further validation consists in testing the model on \emph{check-in} data coming from Foursquare. As we can see in the Methods section, this data set gives us information on the number of visitors who checked in a given venue on the application. 
The dataset contains all Foursquare venues within the M25 motorway, which consists of $300\text{x}10^3$ venues, of which we have filtered just above $100\text{x}10^3$ retail activity (more information on this process is in the Methods section). Each record contains location coordinates, number of check-ins since the venue was registered together with a detailed venue category. For each retailer $r$ in the VOA dataset, we have summed the number of check-ins that were registered in the Foursquare dataset in a range $R$ from the retailer. To choose the range $R$, we have studied the correlation between the Foursquare check-ins and the rent$/m^2$ seen in the VOA dataset. 
\begin{figure}[t]
\centering
\begin{subfigure}[b]{0.48\textwidth}
	\centering
\includegraphics[width=\textwidth,height=6cm]{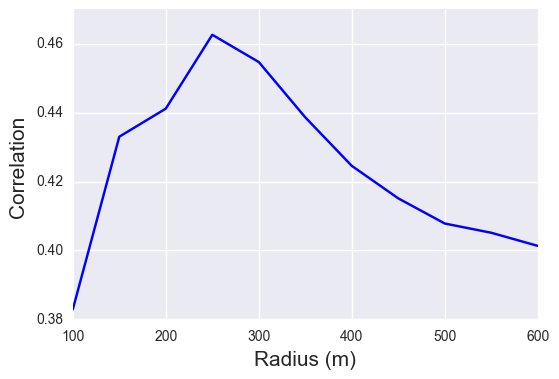}
	\caption{}
	\label{fig:VOA_FOUR}
\end{subfigure}\quad\begin{subfigure}[b]{0.48\textwidth}
	\centering	\includegraphics[width=\textwidth,height=6cm]{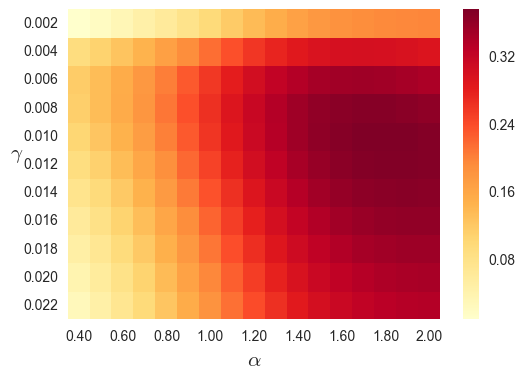}
	\caption{}
	\label{fig:HEAT_FOUR}
\end{subfigure}
\caption{In the left panel in \ref{fig:VOA_FOUR}, we show the correlation between the rateable value of retailers in the VOA dataset,  and the check-ins found in Foursquare around them for different radii. We can see how the correlation peaks at $R =250m$. In the right panel in \ref{fig:HEAT_FOUR} we show the correlation between the expected turnovers $Y(\alpha,\gamma)$ of each shop and the number of check-ins measured within a radius of $R=250m$ from it.  The set of parameters that maximize the correlation are $\alpha_\text{max},\gamma_\text{max} = ( 1.8 , 0.01)$, which are quite similar to what we found for the VOA rateable values.}
\end{figure}
As we can see from fig.
(\ref{fig:VOA_FOUR}) the maximum level of correlation is obtained by fixing $R=250m$ as the range to count check-ins. Curiously enough this is quite close to the agglomeration characteristic exponential decay length. In fig.\ref{fig:HEAT_FOUR} we show the correlation between the modeled turnovers and the check-ins happening in range $R=250m$ from the retailers. We can see how the two parameters for which the correlation is maximum are $(\alpha_\text{max},\gamma_\text{max}) = (1.8, 0.01)$ which, considering the big difference in the data used, can be considered incredibly similar to what we previously obtained. The $\alpha$ value is still in line with a super-linear scaling between floor space and attractiveness, and the $\gamma$ implies a characteristic decay length of $\frac{1}{\gamma}= 100.0 m $, against the $\frac{1}{\gamma}=125m$ previously predicted. Once again given the great distance between the two datasets we have used, we consider that these results are a reassuring sign of the robustness of our findings. 

\begin{figure}[t]
\centering
\begin{subfigure}[b]{0.3\textwidth}
	\centering
	\includegraphics[width=\textwidth,height=\textwidth]{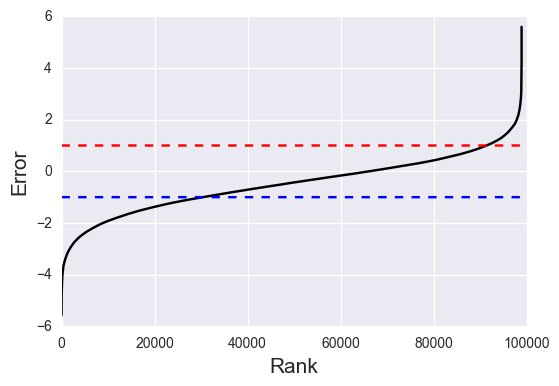}
	\caption{}
	\label{fig:rank}
\end{subfigure}
\quad
\begin{subfigure}[b]{0.3\textwidth}
	\centering
	\includegraphics[width=\textwidth,height=\textwidth]{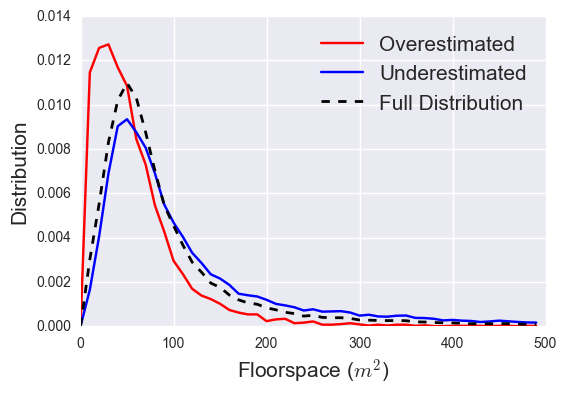}
	\caption{}
	\label{fig:Radius}
\end{subfigure}
\quad
\begin{subfigure}[b]{0.3\textwidth}
	\centering
	\includegraphics[width=\textwidth,height=\textwidth]{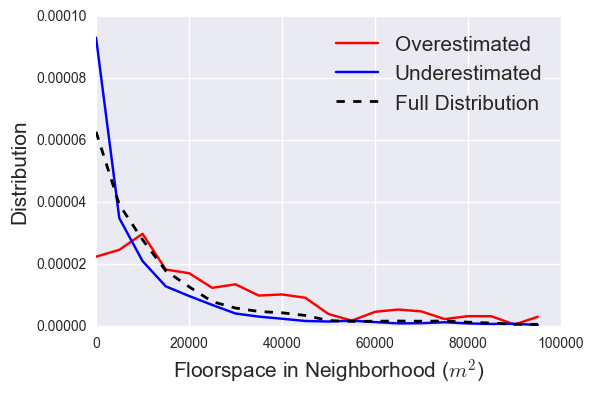}
	\caption{}
	\label{fig:floor}
\end{subfigure}
\caption{In this figure we present the results obtained analyzing the model's errors in estimating the retailers' \emph{success}. In \ref{fig:rank} we have ordered the errors by rank. The red and blue dashed lines indicate the thresholds we have used to define the subset of \emph{overestimated} and \emph{underestimated} retailers. In \ref{fig:Radius} and \ref{fig:floor} we show that the overestimated retailers tend to be smaller than average with more floor space than average in their neighborhood.}
\label{fig:foursquare}
\end{figure}

In order to understand the strength and weaknesses of our model, for each retailer we have studied and compared the difference between the modeled turnover $Y(\alpha_\text{max},\gamma_\text{max})$ and the actual VOA rent. This will allow us to find out for which type of retailers the model is more accurate and which will have to be the future steps to better shape it. The first forced step is to make the two quantities comparable. We call $\mathbf{R}$ the vector of the \emph{fraction} of rents, where the $ith$  component $\frac{R_i}{\sum_iR_{i}}$ represents the fraction of the total rent in the system that belongs to the $ith$ retailer. In the same way, we define the vector of the fraction of modeled turnovers $\mathbf{Y}$, with components $\frac{Y_i}{\sum_i Y_i}$. We can now define the $\emph{E}$ error vector as 
\begin{equation}
\mathbf{E} =\text{log}\left( \frac{\mathbf{Y}}{\mathbf{R}}\right)
\label{eq:e}
\end{equation}
The components of the \emph{error vector} in eq.(\ref{eq:e}) will indicate an overestimation of the \emph{success} of a retailer, if $E_i > 0$ and an underestimation in the opposite case $E_i < 0$.
By observing the characteristics of the retailers for which we underestimate and overestimate the rent, we can see how patterns emerge. In fig.(\ref{fig:rank}) we show a ranked distribution of the errors $E_i$. From the asymmetry of the curve we realize how the model, in this interpretation, tends to underestimate a retailer's success.  Indeed  for the vast majority of the retailers we analysed, we noted that the $Y_i$ is smaller than their $R_i$ which implies that the overestimation errors are typically larger. We can therefore study the characteristics of the retailers we have overestimated, those above the red dashed line, and compare them with those we have underestimated, below the blue line. Observing the red curves in fig.(\ref{fig:Radius}) and fig.(\ref{fig:floor}), we can see how the model tends to overestimate retailers with a \emph{small} floor space surrounded by many neighbors, which looking at our definition of attractiveness in eq.(\ref{eq:A}) is predictable. On the other hand we can see how the blue curves, which represent the distributions of the underestimated retailers are very similar to the the distribution of the full dataset (indicated by the black dashed line).  From the figures it is clear that the model underestimates the number of visits to big retailers in locations with low retail activity concentration, such as supermarkets and other retailers that likely fall under convenience rather than comparison retail, where intra-store variety is more important than inter-store variety. 

\noindent
\newline
\textbf{Outlook.} In this paper we have presented a location choice model, based on random utility theory and following the growing  popularity of the cross-nested choice structure, we have tested this on unconventional data sources of different types. The novelty of the proposed model is that it simulates retailers at the individual level. This opens up exciting opportunities towards integrating the consumer location choice component with explicit retail location micro simulation models able to take full advantage of the emerging availability of detailed data sources while also incorporating complex behavior on price setting, network dynamics and risk management. The proposed model in its current form has been simplified (with no loss of generality) into assuming that all retailers offer unique varieties of the same product. Moreover, it has been assumed that (i) all consumers have equal disposable retail budgets regardless of their location, (ii) all trips are uni-purpose (only shopping is considered), (iii) VOA rateable values are good indicators of floor space rents and (iv) product prices do not vary in space.  These assumptions mean that a considerable part of the complexity of the decision making mechanism is not represented by the model. The VOA and LTDS datasets that we are currently using have the potential to increase the complexity of the model significantly towards removing some of the existing simplifying assumptions, and when combined with passively collected social media datasets and formal datasets on economic activity (e.g. from the Business Structure Dataset developed by ONS), these models could offer sufficient detail to capture all the main dimensions of behavioral variation. Having said this, the basic model that we present in this paper remains very useful both as the baseline example of the proposed approach and as a benchmark; despite its simplicity, it is able to capture and reproduce both internal and external economies of scale very closely. Further immediate lines of research could consider different categories of retailers separately as well as applying the same analysis on data coming from other cities. Looking at the not too distant future, passively generated datasets of human presence promise to offer deeper insights into the dynamics of urban activities, including spatio-temporal patterns of shopping behavior. Having said that, the abundance of existing social media datasources and the relentless pace in which new data are currently being introduced, sustain the promise of accessible and highly dis-aggregated spatiotemporal information for anyone who manages to overcome lack of specification, representational biases and possibly absence of context.

\section*{Methods}

\section{Datasets}

In this section, we will trawl the datasets we have used to calibrate and validate the model. These are a combination of formal proprietary datasets of travel behavior and economic activity, and passively collected data sources of digital social media footprints.

\noindent
\newline
\textbf {London Travel Demand Survey.} LTDS is a continuous household survey of the London area, covering the London boroughs as well as the area outside Greater London but within the M25 motorway. Results in the most recent report relate to residents of the Greater London area, comprising the 32 London boroughs and the City of London. The first year of results covered the financial year 2005/06, meaning that there are now eight years of data available.
The survey is a successor to the household survey component of the London Area Transport Survey (LATS) which was last carried out in 2001. The LTDS annual sample size is around 8,000 households in a typical year, a sum of 65,000 households for the 2005-2013 period.
LTDS captures information on households, people, trips and vehicles. All members of the household are surveyed, with complete trip detail for a single day recorded for all household members aged 5 and over. Three questionnaires are used for the household questionnaire, individual questionnaires for all household members, and trip sheets for travel diaries. The latter capture data on all trips made on a designated travel day, the same day for all members of the household. Details captured include trip purposes, modes used, trip start and end times, and the locations of trip origins and destinations. We have used this dataset only to calibrate the parameter in the cost function in eq.(\ref{eq:cost}).

\noindent
\newline
\textbf{Valuation Office Business Rates.}
Since, the recent online publication of the Valuation Office Agency Business Rates for 2005 and 2010, the business rates of all business premises in England and Wales have become available to the public and offer a unique in-depth extent and geographic precision dataset. VOA compiles and maintains lists of rateable values of the 1.7 million non-domestic properties in England, and the 100,000 in Wales, to support the collection of around £25 billion in business rates.
The Rateable value represents the Agency's estimate of the open market annual rental value of a business/ non-domestic property; i.e. the rent the property would let for on the valuation date, if it were being offered on the open market.
The rateable value is estimated based on the varying rents at the vicinity of a property. This represents a reasonable level of open market rental value, taking into account the expected turnover of the premise, the size, age and condition of the property, the length of the frontage, the depth and vertical layout, the visibility, the footfall and pedestrian flow volumes on surrounding streets etc. Because the rateable value of a property is used to determine the non-domestic property tax (business rate), and the evaluators have access to detailed information (such as contracts, revenue documents etc.), and because of this comprehensive evaluation documentation and methodology, rateable value is considered a very good indicator of the property value of a hereditament.
The agency publishes a detailed set of information for each property; this includes the classification of its main use (detailed breakdown into more than 100 classes), the full address and postcode, the total area of the premise, the total rateable value and breakdown into zones with different rateable value per square metre, and the weighted average rateable value per square metre. This makes it possible to create a detailed map of rateable value for any use. {\color{black}. As said the dataset contains information on more than $3\text{x}10^5$ non-domestic activities in the area of London. These are labeled with a \emph{scat code} which depends of the category they belong to. For this analysis we have selected only the ones which belong to retailing scat codes, which are shown in table \ref{table}}.

\begin{table}[t]
\centering
\begin{tabular}{l|r}
Category Description& Scat Code \\\hline
SHOPS & 249 , 155 , 152 , 210 , 154 \\
HAIRDRESSING SALONS  & 1910 ,  23 \\
SHOWROOMS & 251 , 249  \\
KIOKS & 243 , 249 \\
CAR SHOWROOMS & 42 \\
RETAIL WAREHOUSES & 235 \\
SUPERSTORES & 139  , 152 \\
MARKETS & 165
\end{tabular}
\caption{\label{tab:widgets} In this table we show the categories we have selected in the VOA datasets with their relative Scat Codes.}
\label{table}
\end{table}

\noindent
\newline
\textbf{Social Media Spatiotemporal Profiles: Foursquare.}
{\color{black} We have used one passively generated dataset. The dataset contains all Foursquare venues within the M25 motorway (300 thousands venues). Each record contains location coordinates, number of check-ins and unique visitors since venue was registered and detailed venue category (activity type). From all the entries in the Dataset we have filtered those belonging to the same categories in table \ref{table}. The Foursquare venue data was collected in December 2014.}

\section*{Acknowledgements }

{\color{black} All authors thank the EU FP7 for their support through the INSIGHT Project (Innovative Policy Modelling and Governance Tools for Sustainable Post-Crisis Urban Development) (http://www.insight-fp7.eu). M. B. wishes to thank the ERC for support through grant no. 249393-ERC-2009-AdG. D.P. is grateful to C. Molinero and R. Morphet for continuous and useful discussions, and to A. Ialongo for a long and fruitful conversation on the analysis of the results. Authors would also like to acknowledge the contribution of C. Vargas-Ruiz, J. Serras and P. Ferguson in the development of the the model as described in \cite{Zachariadis2015}

\section*{Author contributions statement}
{ \color{black} DP and VZ contributed equally to the scientific research in this work. VZ developed the model, DP implemented it and performed the  simulations, and they both analysed the results. MB directed the project as PI. All authors were involved in the writing of the paper.}


\begin{thebibliography}{10}

\bibitem{batty1995new}
M.~Batty.
\newblock New ways of looking at cities.
\newblock {\em Nature}, 377:574, 1995.

\bibitem{bettencourt2010unified}
L.~Bettencourt and G.~West.
\newblock A unified theory of urban living.
\newblock {\em Nature}, 467(7318):912--913, 2010.


\bibitem{crucitti2006centrality}
P.~Crucitti, V.~Latora, and S.~Porta.
\newblock Centrality measures in spatial networks of urban streets.
\newblock {\em Physical Review E}, 73(3):036125, 2006.


\bibitem{barthelemy2011spatial}
M.~Barth{\'e}lemy.
\newblock Spatial networks.
\newblock {\em Physics Reports}, 499(1):1--101, 2011.


\bibitem{batty2013big}
M.~Batty.
\newblock Big data, smart cities and city planning.
\newblock {\em Dialogues in Human Geography}, 3(3):274--279, 2013.


\bibitem{louf2013modeling}
R.~Louf and M.~Barthelemy.
\newblock Modeling the polycentric transition of cities.
\newblock {\em Physical Review Letters}, 111(19):198702, 2013.

\bibitem{louf2014congestion}
R.~Louf and M.~Barthelemy.
\newblock How congestion shapes cities: from mobility patterns to scaling.
\newblock {\em Scientific Reports}, 4, 2014.

\bibitem{huff1966programmed}
D.~L. Huff.
\newblock A programmed solution for approximating an optimum retail location.
\newblock {\em Land Economics}, 42(3):293--303, 1966.

\bibitem{wilson1969use}
A.~G. Wilson.
\newblock The use of entropy maximising models, in the theory of trip
  distribution, mode split and route split.
\newblock {\em Journal of Transport Economics and Policy}, pages 108--126,
  1969.

\bibitem{wilson2011entropy}
A.~G. Wilson.
\newblock Entropy in urban and regional modelling, 2011.



\bibitem{manchanda1999shopping}
P.~Manchanda, A.~Ansari, and S.~Gupta.
\newblock The shopping basket: A model for multicategory purchase
  incidence decisions.
\newblock {\em Marketing Science}, 18(2):95--114, 1999.

\bibitem{Teller2008}
C.~Teller and T.~Reutterer.
\newblock {The evolving concept of retail attractiveness: What makes retail
  agglomerations attractive when customers shop at them?}
\newblock {\em Journal of Retailing and Consumer Services}, 15(3):127--143,
  2008.

\bibitem{williams1977formation}
H.~C. Williams.
\newblock On the formation of travel demand models and economic evaluation
  measures of user benefit.
\newblock {\em Environment and Planning A}, 9(3):285--344, 1977.

\bibitem{Heppenstall2013}
A.~J. Heppenstall, K.~Harland, A.~N. Ross, and D.~Olner.
\newblock {Simulating spatial dynamics and processes in a retail gasoline
  market: An agent-based modeling approach}.
\newblock {\em Transactions in GIS}, 17(5):661--682, 2013.

\bibitem{vanhaverbeke2011agent}
L.~Vanhaverbeke and C.~Macharis.
\newblock An agent-based model of consumer mobility in a retail environment.
\newblock {\em Procedia-Social and Behavioral Sciences}, 20:186--196, 2011.


\bibitem{hotelling1990stability}
H.~Hotelling.
\newblock Stability in competition.
\newblock In {\em The Collected Economics Articles of Harold Hotelling}, pages
  50--63. Springer, 1990.



\bibitem{von1966isolated}
J.~H. von Th{\"u}nen and P.~G. Hall.
\newblock {\em Isolated state: an English edition of Der isolierte Staat}.
\newblock Pergamon, 1966.

\bibitem{christaller1966central}
W.~Christaller.
\newblock {\em Central places in southern Germany}.
\newblock Prentice-Hall, 1966.

\bibitem{dennis2002central}
C.~Dennis, D.~Marsland, and T.~Cockett.
\newblock Central place practice: shopping centre attractiveness measures,
  hinterland boundaries and the uk retail hierarchy.
\newblock {\em Journal of Retailing and Consumer Services}, 9(4):185--199,
  2002.
  
  
\bibitem{alonso1960theory}
W.~Alonso.
\newblock A theory of the urban land market.
\newblock {\em Papers in Regional Science}, 6(1):149--157, 1960.



\bibitem{krugman1990increasing}
P.~Krugman.
\newblock Increasing returns and economic geography.
\newblock Technical report, National Bureau of Economic Research, 1990.



\bibitem{krugman1998s}
P.~Krugman.
\newblock What's new about the new economic geography?
\newblock {\em Oxford Review of Economic Policy}, 14(2):7--17, 1998.


\bibitem{dixit1977monopolistic}
A.~K. Dixit and J.~E. Stiglitz.
\newblock Monopolistic competition and optimum product diversity.
\newblock {\em The American Economic Review}, 67(3):297--308, 1977.



\bibitem{fiasconaro2016spatio}
A.~Fiasconaro, E.~Strano, V.~Nicosia, S.~Porta, and V.~Latora.
\newblock Spatio-temporal analysis of micro economic activities in rome reveals
  patterns of mixed-use urban evolution.
\newblock {\em PloS One}, 11(3):e0151681, 2016.

\bibitem{suarez2004accounting}
A.~Suarez, I.~R. del Bosque, J.~M. Rodriguez-Poo, and I.~Moral.
\newblock Accounting for heterogeneity in shopping centre choice models.
\newblock {\em Journal of Retailing and Consumer Services}, 11(2):119--129,
  2004.
  
  
\bibitem{anderson1992discrete}
S.~P. Anderson, A.~De~Palma, and J.~F. Thisse.
\newblock {\em Discrete Choice Theory of Product Differentiation}.
\newblock MIT press, 1992.

\bibitem{train2009discrete}
K.~E. Train.
\newblock {\em Discrete choice methods with simulation}.
\newblock Cambridge university press, 2009.



\bibitem{daly1978improved}
A.~Daly and S.~Zachary.
\newblock Improved multiple choice models.
\newblock {\em Determinants of travel choice}, 335:357, 1978.


\bibitem{ben1985discrete}
M.~E. Ben-Akiva and S.~R. Lerman.
\newblock {\em Discrete choice Analysis: Theory and Application to Travel
  Demand}, volume~9.
\newblock MIT press, 1985.

\bibitem{bierlaire2006theoretical}
M.~Bierlaire.
\newblock A theoretical analysis of the cross-nested logit model.
\newblock {\em Annals of Operations Research}, 144(1):287--300, 2006.


\bibitem{wen2001generalized}
C.-H. Wen and F.~S. Koppelman.
\newblock The generalized nested logit model.
\newblock {\em Transportation Research Part B: Methodological}, 35(7):627--641,
  2001.


\bibitem{mcfadden2001economic}
D.~McFadden.
\newblock Economic choices.
\newblock {\em The American Economic Review}, 91(3):351--378, 2001.




\bibitem{mcfadden1980econometric}
D.~McFadden.
\newblock Econometric models for probabilistic choice among products.
\newblock {\em Journal of Business}, pages S13--S29, 1980.


\bibitem{millward2013active}
H.~Millward, J.~Spinney, and D.~Scott.
\newblock Active-transport walking behavior: destinations, durations,
  distances.
\newblock {\em Journal of Transport Geography}, 28:101--110, 2013.




\bibitem{yang2012walking}
Y.~Yang and A.~V. Diez-Roux.
\newblock Walking distance by trip purpose and population subgroups.
\newblock {\em American journal of Preventive Medicine}, 43(1):11--19, 2012.

\bibitem{Zachariadis2015}
V.~Zachariadis, C.~Vargas-ruiz, J.~Serras, and P.~Ferguson.
\newblock {Decoding Retail Location : A Primer for the Age of Big Data and
  Social Media}.
\newblock {\em Cupum 2015}, 2015.

\end{thebibliography}
\end{document}